# Observation of hidden nuclear reactions on fast neutron-irradiated Lu isotopes

**Ihor Kadenko[1,2] · Barna Biró[3] · András Fenyvesi[3] · Vladyslav Morozyuk[2] · Kateryna Okopna[2] · Nadiia Sakhno[1]**

**Abstract**
Some nuclear reaction channels may not be easily identified, but they can still contribute to very important characteristics like nuclear reaction cross-sections, isomeric yields, etc. The dineutron, as a bound nucleus of two neutrons, can play hidden roles, as it is considered a product resulting from fast neutron-induced nuclear reactions involving the isotopes of lutetium. When being formed in the output channels, a bound dineutron allows for an explanation for unexpected enhancement of reaction cross-sections. We used available data and instrumental spectra to deeply consider the role of the dineutron in order to understand the consequences of its formation and subsequent decay upon transformations of nuclear reaction products. As a result, we obtained experimental confirmation that the half-life of $^{176g}$Lu is significantly diminished within a certain time interval, which reveals an extremely high estimation for its cross-section of ~$10^{11}$ b. This is actually due to $^{176g}$Lu "burnup," followed by the fusion between $^{175}$Lu and a deuteron. This introduces the existence of novel, low-energy reactions and fusion pathways under neutron irradiation, which in-turn, leads to implications for both basic nuclear physics and high-energy physics, along with their applications.

## 1 Introduction

In order to study specific nuclear reaction cross-section(s), isomer ratio(s), and possible product(s) in the outgoing channels, it is common practice to first develop an idea and then designing and conducting an experiment using the activation technique to test the idea. The steps of the experiment are:

1) selection of a proper irradiation facility;
2) choosing the proper activation sample and optimizing it and their radiation geometry,
3) irradiation of the sample,
4) after the irradiation and some cooling time repeated counting of the induced radioactivity of the induced activity of the sample.

Step 4 is needed to reduce the possible statistical and systematic uncertainties. Once the raw data is obtained, the necessary calculations of reaction cross-sections etc. can be made while considering the experimental details needed for finalizing and drafting the results. Even after all of this, there will come the moment when those results will probably be reconsidered or even criticized based on new (nuclear) data, knowledge, or other research discoveries.

We tried to apply the approach, outlined in steps 1-4, while reconsidering some previously published studies that appear to contain discrepancies or unclear manifestations, while taking into account more modern assumptions and new experimental data.

Qaim, et al. [1] found that the cross-section of the $^{127}$I($n$, 3$n$)$^{125}$I nuclear reaction equals 40 ± 15 mb when measured using 14.7 MeV neutrons, while the threshold energy is 16.42 MeV for this nuclear reaction. Hankle et al. [2] found that the nuclear reaction cross-section of $^{197}$Au($n$,$^3n$)$^{195}$Au was 61 ± 20 mb using 14.4±0.4 MeV neutrons. They stated "…it arises only from the high-energy tail of the neutron distribution…" because the threshold for this reaction is 14.79 MeV [3]. However, TENDL [4] evaluates only 49 $\mu$b cross-section at 15 MeV neutron energy, while ~ 60 mb cross-section is listed at about 16.25 MeV neutron energy for this reaction. All of these means that another (hidden) nuclear reaction may take place at 14.4±0.4 MeV neutron energy. Then a possible explanation would be the emission of a bound nucleus consisting of two neutrons, and the emission of an additional neutron in the outgoing channel. It has

The research carried out at HUN-REN ATOMKI was supported by the TKP2021-NKTA-42 project financed by the National Research, Development and Innovation Fund of the Ministry for Innovation and Technology, Hungary. The MGC-20 cyclotron of HUN-REN ATOMKI is a Research Infrastructure of the Cluster of Low Energy Accelerators for Research (CLEAR) of the EURO-LABS project. I. Kadenko and N. Sakhno were supported by the Transnational Access of the CLEAR EURO-LABS project. The EURO-LABS project has received funding from the European Union's Horizon Europe Research and Innovation programme under Grant Agreement No. 101057511.

✉ Ihor M. Kadenko
imkadenko@univ.kiev.ua

[1] International Nuclear Safety Center of Ukraine of Taras Shevchenko National University of Kyiv, Kyiv, 03022, Ukraine

[2] Department of Nuclear and High Energy Physics, Faculty of Physics, Taras Shevchenko National University of Kyiv, Kyiv, 01601, Ukraine

[3] HUN-REN Institute for Nuclear Research (HUN-REN ATOMKI), Debrecen, 4032, Hungary

to emphasize that the dineutron cannot be directly observed because it decays instantaneously after its escape. However, the radioisotope product of the nuclear process can be observed via detecting its decay radiations.

Since the energy conservation law is not the subject to doubt, we can formulate the assumption for neutron-irradiated $^{175}$Lu reactions [5,8], which involve the formation of bound dineutrons with binding energies between 1.7 MeV [5] and 2.8 MeV [6]. Both of these reactions, along with those from earlier studies, involve the target nuclei $^{159}$Tb [6] and $^{197}$Au [7]. These studies were extended by including the $^{175}$Lu($n$, $^{2}n$)$^{174g}$Lu nuclear reaction [5]. Therefore, the subject of this study covers thorough re-considerations of our previous experimental results [5,8] obtained from fast neutron irradiations of Lu samples of natural isotopic abundance, promoting the search for hidden nuclear reactions and their unexpected manifestations.

## 2 Experimental setups

Neutrons, in the range of 5 - 6 MeV, were generated by bombarding a $D_2$-gas target with deuterons produced at the MGC-20E cyclotron located in the HUN-REN ATOMKI facility. The resulting quasi-monoenergetic neutrons, produced via the $^{2}$H($d$, $n$)$^{3}$He nuclear reaction, were then used to irradiate the Lu samples. The details of the irradiation experiment are given in [5] and [7]. Lu irradiation with 14.6 MeV neutrons was carried out by using the D-T neutron generator, NG-300, located at the Department of Nuclear and High Energy Physics of Taras Shevchenko National University of Kyiv. More details are available in our earlier published papers [8-10]. After completion of the irradiations, the Lu samples were counted on an HPGe spectrometer.

## 3 Data analysis and results

We have been using the same spectrum obtained from the Lu-1 +Lu-2 1st samples described in [5] along with another spectrum of Lu samples Lu-1 +Lu-2 BS counted at HUN-REN ATOMKI prior to neutron irradiation. Both spectra are shown in Fig.1 with the corresponding data listed in Table 1.

**Table 1** Summary of the gamma counting measurements performed before and after the irradiation stopped at 2023.05.19 17:00:00 (Detector: MIRION CANBERRA HPGe XtRa)

| Measurement ID | START | Live time (s) | Real time (s) |
|---|---|---|---|
| Lu-1 +Lu-2 BS | 5/18/2023 12:50 PM | 4,762.5 | 4,773.1 |
| Lu-1 +Lu-2 1st | 5/22/2023 7:23 AM | 713,893.3 | 715,246.1 |

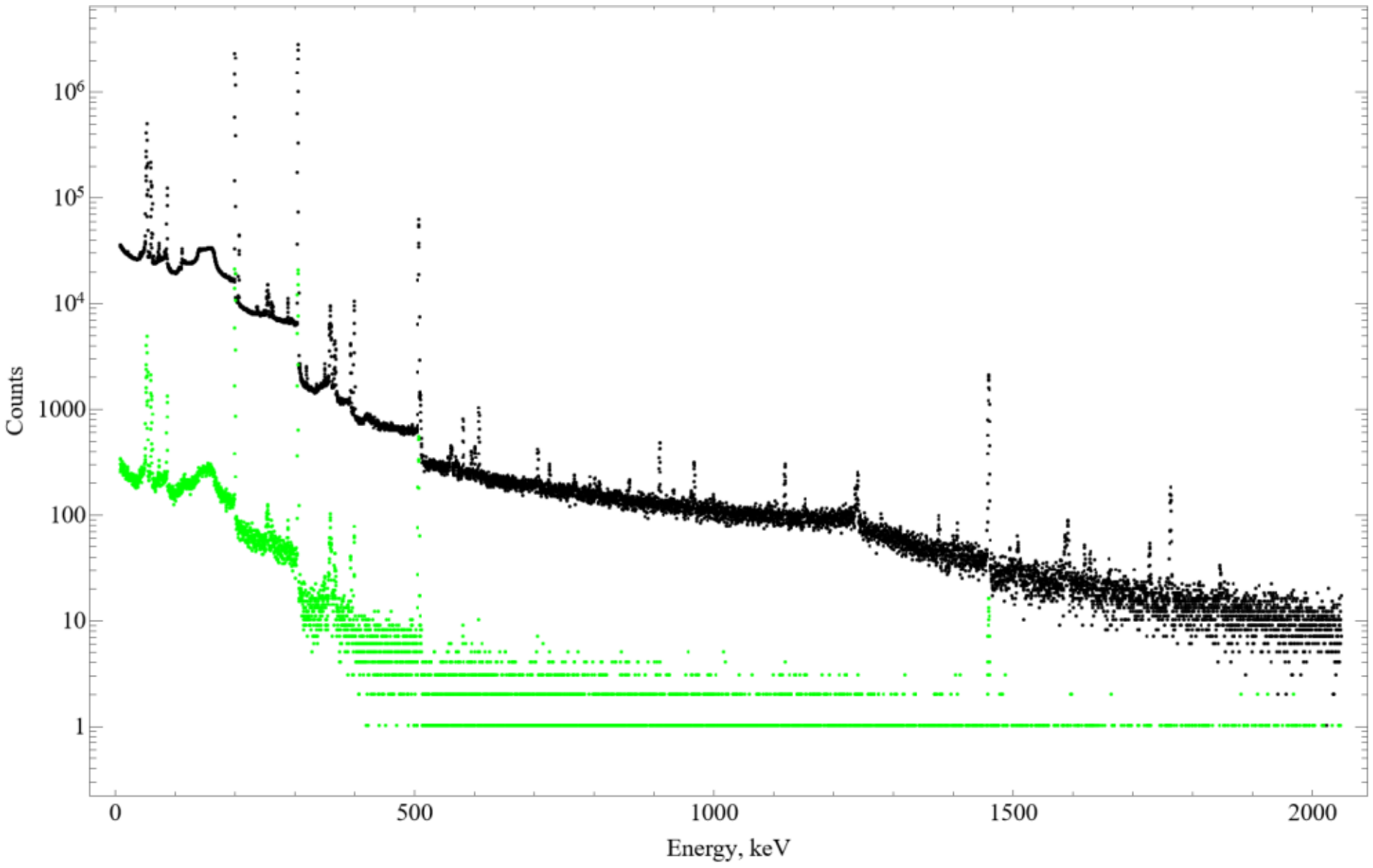

**Fig.1** General view of the spectrum 1st (black) and the background spectrum BS (green) with Lu samples

### 3.1 The nuclear reaction $^{176}$Lu($n$, $\gamma$)$^{177m,g}$Lu cross-section measurement

In [8], the cross-section of the $^{176}$Lu($n$, $\gamma$)$^{177m,g}$Lu nuclear reaction, at 14.6 MeV neutron energy, was measured for the first time. The empirical results are of a significantly greater value than those obtained from the TALYS-1.96 and Empire 3.2.3 codes or from the TENDL and ENDF databases. A very thorough analysis of the scattered neutron's contribution can only account for a very small portion of the discrepancy, which is not sufficient to explain the existing difference between the measured and calculated values. It is important to remember that the decay of $^{177m,g}$Lu actually leads to gamma-rays irradiated by the $^{177}$Hf isotope. Therefore, if we could find another way to populate the levels of $^{177}$Hf, it would serve as an explanation, because it would suggest that there is another reaction channel open: $^{176}$Lu($n$,$^{2}n$)$^{175}$Lu. In this reaction, a bound dineutron can be formed near the surface of $^{175}$Lu, that is within its potential well [5]. Then the dineutron undergoes the $\beta^-$ - decay and transforms into a deuteron. Subsequent fusion of this deuteron with the $^{175g}$Lu nucleus occurs: $^{175}$Lu($d$, $\gamma$)$^{177}$Hf. This transformation chain, including the conversion of the dineutron into the deuteron, was discussed earlier [11], and it led us to the conclusion that an excited state of $^{177}$Hf was formed. It is the gamma irradiation due to the prompt relaxation of the $^{177*}$Hf that was previously detected and assigned to the decay of $^{177m,g}$Lu.

### 3.2 Remarks on the cross sections measured previously for the $^{175}$Lu(n, $^{2}$n)$^{174g}$Lu nuclear reaction

In [8], we did not notice the implications of the $^{175}$Lu($n$, $^{2}n$)$^{174}$Lu cross section measurements, but applying the same technique as described in [8], for the 1,241.8 keV peak with 5,805 ± 89 counts, we get the following estimate of its reaction cross-section: 2.54 ± 0.23 b. Taking into account the TENDL [4] value of 2 b for this neutron energy region, we have an excess of ~ 0.54 ± 0.23 b. It is important to note that this reaction cross-section value also includes another contribution from the $^{176}$Lu($n$, $^{3}n$)$^{174}$Lu open nuclear reaction channel [12], for which no experimental data is available [13]. Again, TENDL [4] yields a ≈ 20 mb cross-section for the $^{176}$Lu($n$, $^{3}n$)$^{174}$Lu reaction cross-section for this neutron energy region.

### 3.3 Estimation for the cross section of the $^{176}$Lu($n$,$^{2}n$)$^{175}$Lu nuclear reaction

In our earlier work [5], we reported the formation of a bound dineutron in the $^{175}$Lu($n$, $^{2}n$)$^{174g}$Lu nuclear reaction and estimated the cross-section at ~ 6 MeV neutron energy. Comparison of this ($n$, $^{2}n$) reaction to the similar one for $^{176}$Lu (2.599% natural abundance), while using the criteria approach published in [14], it can be concluded that the second reaction is not favored. However, having the necessary spectra for detailed processing, we examined the possibility of providing another evidence for the formation of a bound dineutron as well. Unlike the case of the $^{175}$Lu($n$, $^{2}n$)$^{174g}$Lu nuclear reaction, a stable product exists in the outgoing channel of the $^{176}$Lu($n$, $^{2}n$)$^{175}$Lu reaction. Hence, an emission cannot be observed with the off-line gamma-spectrometry, which would be needed to prove the formation of a bound dineutron. Therefore, we applied another approach suggested by our co-authors K.O. and V.M. Using the spectrum for the Lu-1 +Lu-2 BS, measured for the background, and the Lu-1 +Lu-2 1st spectrum [5], measured for the lutetium samples (see Table 1 and Table 2), we calculated and followed the time dependence of the net count rates for the 88.3 keV, 201.8 keV, and 306.8 keV gamma photons emitted during the de-excitation of $^{176*}$Hf, which is the product of the $\beta^-$ -decay of $^{176g}$Lu. The decay scheme of $^{176g}$Lu is shown in Fig. 2.

The results are given in Table 2, where the first line includes the energies and intensities for the most intense gamma-transitions; the last row contains the ratios of the peak count rates with their corresponding uncertainties. One must emphasize that all three count rates are significantly reduced, even though both spectrum acquisitions took place under identical counting geometries. This means that there is either an essential decrease in the amount of $^{176g}$Lu nuclei after irradiation in a fast neutron field, and/or there is a breakdown of the $^{176g}$Lu decay scheme. In the last two columns of Table 2, the count rates of the true coincidence sum peaks at 508.6 (306.8+201.8) and 290.1 (201.8+88.3) keV energies showed a similar decrease. Such reductions can be quantified and characterized in different ways.

The background corrected net count rates, presented in Table 2, were used to estimate of the half-life of $^{176g}$Lu decay after fast neutron irradiation. The following equation was used:

$$T_{1/2} = \ln 2 \times \Delta t / \ln\left(I_{BS} / I_0\right), \qquad (1)$$

where $\Delta t$ is the time transpired between the acquisitions of the spectra. The peak count rates for $I_{BS}$ and $I_0$, are shown in Table 1. The peak count rates for the Lu-1 +Lu-2 BS and Lu-1 +Lu-2 1st spectra were carried using the same peak energies.

We used the 306.8 keV peak as a reference for this estimate and assumed a negligible decay of $^{176g}$Lu during that time interval ($\Delta t$). Taking the data from Tables 1 and 2, we get the following estimate: $T_{1/2}$ = 62.2±0.2 d.

**Table 2** Count rates measured for the peaks at 88.3 keV, 201.8 keV and 306.8 keV and for the corresponding sum peaks at 290.1 keV and 508.6 keV gamma energies. (Detector: MIRION CANBERRA HPGe XtRa)

| Measurement ID | $E_\gamma$ ($I_\gamma$) 88.3 keV (14.5%) | 201.8 keV (78%) | 306.8 keV (93.6%) | 508.6 keV | 290.1 keV |
|---|---|---|---|---|---|
| | Peak count rate (1/s) | | | | |
| Lu-1 +Lu-2 BS | 0.74 ± 0.02 | 15.83 ± 0.06 | 17.56 ± 0.06 | 0.563 ± 0.012 | 0.032 ± 0.009 |
| Lu-1 +Lu-2 1st | 0.423 ± 0.001 | 11.243 ± 0.004 | 15.324 ± 0.005 | 0.403 ± 0.001 | 0.0230 ± 0.0008 |
| Ratio | 1.75 ± 0.02 | 1.41 ± 0.06 | 1.15 ± 0.06 | 1.397 ± 0.012 | 1.391 ± 0.009 |

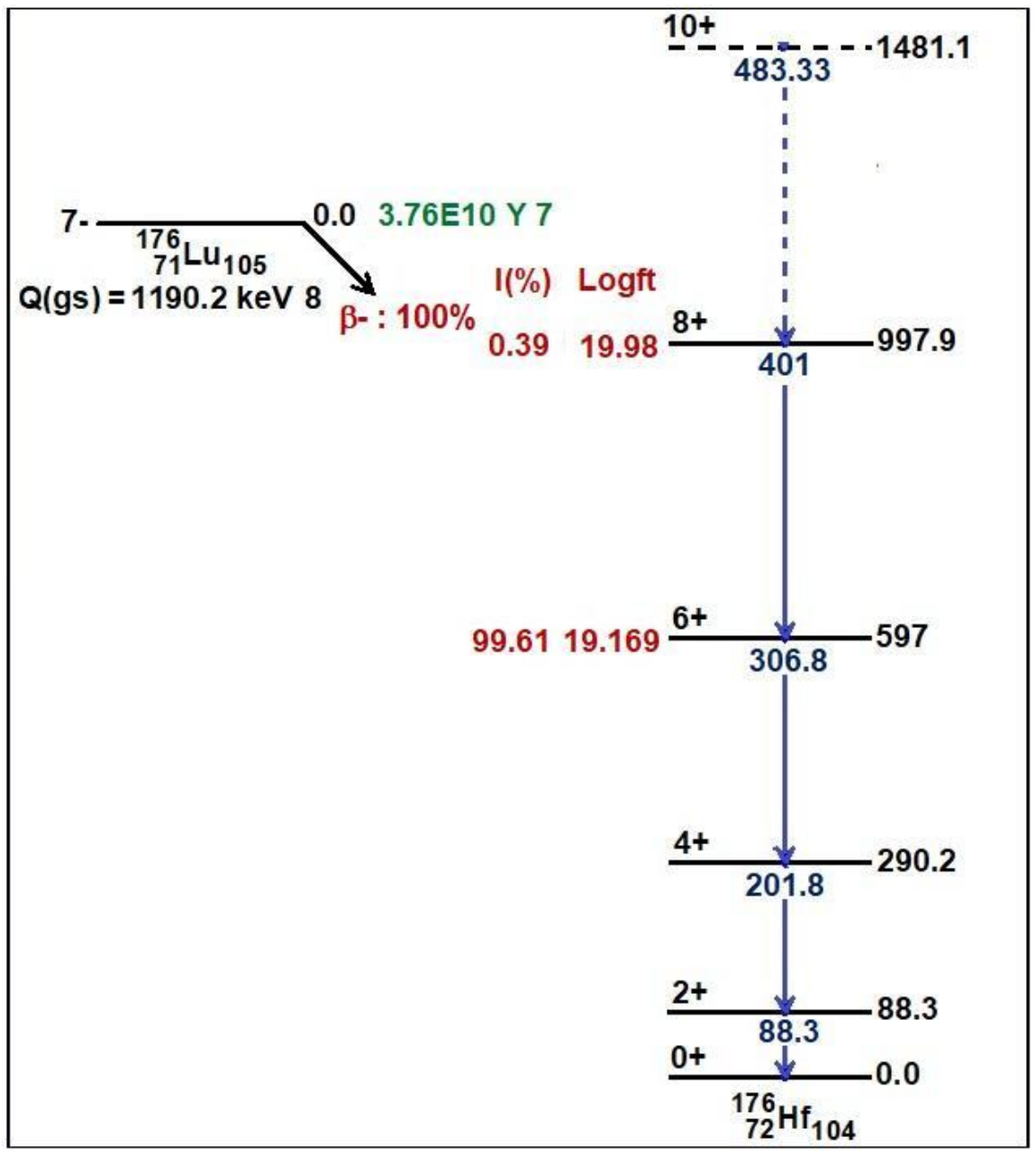

**Fig. 2** Decay scheme of $^{176g}$Lu [15]

This value dramatically differs from the well-known half-life of $^{176g}$Lu: 3.76×10$^{10}$ y [15]. It means that we faced a typical burnup problem, often addressed during nuclear power production [16]. The following expression can be used for the description of this process:

$$I_0 = I_{BS} \times EXP(-\sigma \times \varphi \times t_{irr}), \qquad (2)$$

where $\varphi$ is the flux of the fast neutrons (cm$^{-2}$s$^{-1}$) and $t_{irr}$ is the duration of the neutron irradiation (s) [5]. From Eq. 2 we obtain $\sigma$ = 3.88±0.12×10$^{-13}$ cm$^2$, which corresponds to 3.88±0.12×10$^{11}$ b! As to our knowledge, no known nuclear process comes within orders of magnitude of this reaction cross-section.

### 3.4 The nuclear reactions $^{174,175}$Lu(*d*, *γ*)$^{176,177}$Hf

In our paper [11], we tried to mathematically describe the process of fusion between a heavy nucleus and a deuteron resulting from the decay of a bound dineutron into a deuteron, an electron and the anti-electron neutrino [5], while it is located within the potential well of the heavy nucleus. This phenomenon takes place at ambient temperatures. Being within the range of the strong interaction, such heavy and light nuclei may indeed fuse, forming a corresponding product nucleus in the outgoing channel. Such a nuclear process does not require high temperatures for fusion to occur [17]. In an attempt to detect the decay of the product resulting from the fusion between Lu isotope nuclei and the deuteron, we consider the formation of $^{176,177m,g}$Lu and $^{177}$Hf.

Accordingly, to prove the occurrence of such fusion phenomenon, we need to account for the following fact: the decrease in the number of heavy nuclei ($^{175,176}$Lu) in the output channel should result in an increase in the atomic number of the product nuclei. This is the result of fusion converting Lu to Hf. We can detect an increase in the number of such nuclei by measuring the increased activity of gamma quanta inherent in their relaxation/decay. However, we faced the fact that there is a large amount of $^{175}$Lu, in natural abundance in the original sample. Thus, it is not easy to record a decrease in the number of nuclei of this isotope. Therefore, the only source of information will be the $^{176}$Lu/$^{176}$Hf ratio. $^{176}$Hf is a decay product of $^{176g}$Lu, therefore the lowest energy decay lines in them overlap. Gamma-transitions below the excitation level of 1,481 keV, for example the 483.3 keV line, will be a characteristic of $^{176}$Hf (Fig. 2). As for the case of $^{176}$Hf, we observed all four gamma-lines due to decay of $^{176g}$Lu and did not observe the gamma-transition of 483.33 keV (Fig. 2). Such a combination of observable gamma-lines could be explained by either no occurrence of the fusion reaction ($^{174}$Lu(*d*, *γ*)$^{176}$Hf) at all, or by the fact that the excitation energy of the $^{176}$Hf nucleus is lesser than 1,481.1 keV. We will keep this fact in mind and will get back to this issue below. The amount of $^{176}$Lu, on the other hand, decreased sharply after neutron irradiation, and does not allow for an additional accumulation of $^{176}$Lu nuclei. Instead, a decrease in the number of $^{176}$Lu nuclei may denote the above-described process, however, for a different reaction, namely the $^{176}$Lu(*n*, $^2$*n*)$^{175}$Lu nuclear reaction.

Let us now consider the second fusion reaction: $^{175}$Lu(*d*, *γ*)$^{177}$Hf. Again, we remember the decay scheme of $^{177}$Hf (Fig. 3) and look for gamma-peaks in the instrumental gamma spectrum (Fig. 4), matching this decay scheme. The most obvious way for $^{177m,g}$Lu to decay into $^{177}$Hf is the following reaction: $^{176}$Lu($n$, $\gamma$)$^{177m,g}$Lu. We do not need to consider an isomeric state of $^{177}$Lu with its 6 $\mu$s half-life. Another isomeric state $^{177m}$Lu decays with the 160.4 d half-life and emits gamma photons with the corresponding intensities [15]: 112.9 keV (21.4%); 208.4 keV (55.4%); 228.5 keV (35.9%); 378.5 keV (29.4%) and 418.5 keV (21.7%) that would necessarily be easily detectable in our experiment.

However, in our Lu-1+Lu-2 1$^{st}$ spectrum we clearly observe only the first two emissions with 36,060 ± 457 counts and 128,100 ± 442 counts, respectively. The last three peaks did not show up in this spectrum. This means that the isomeric state of $^{177m}$Lu with a 970.2 keV energy level is not there, and the presence of the first two gamma peaks can be explained by the decay of ground state of $^{177g}$Lu. The 177$g$ isotope of Lu has a half-life 6.64 d; its decay can be followed by the counting of the following gammas: 112.9 keV (6.2%) and 208.4 keV (10.4%). The highest energy level of this decay chain is only 321.3 keV. Keeping in mind our assumption about the possible fusion reactions, $^{174g}$Lu($d$, $\gamma$)$^{176}$Hf and $^{175}$Lu($d$, $\gamma$)$^{177}$Hf, we now look for direct gamma transitions in the $^{176}$Hf and $^{177}$Hf decay schemes.

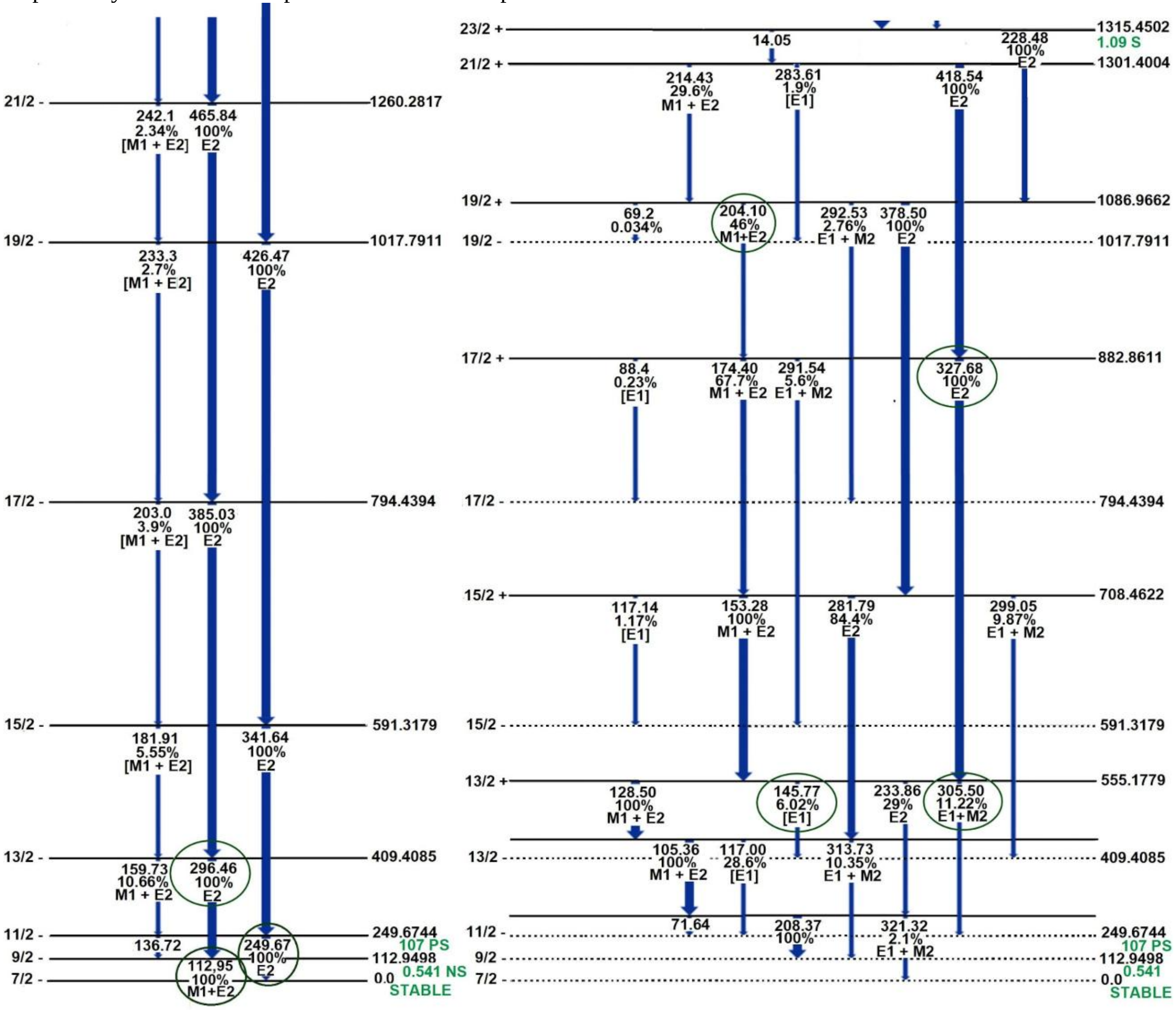

**Fig. 3** Partial decay scheme of $^{177}$Hf [15]

From Fig. 3 and Fig. 4, one can conclude that fusion takes place between $^{175}$Lu and the deuteron, resulting in the excited state of $^{177}$Hf. Since neither an isomeric state of this nucleus with the half-life of 1.09 s and the gamma-transition of 228.48 keV is observed, nor is a gamma-ray transition at 418.5 keV present in the instrumental spectrum, the excitation energy of $^{177}$Hf must be greater than 1,086.97 keV and less than 1,301.4 keV. This is true even though the 204.1 keV transition is present. We can now use the following equation to estimate the excitation energy for this nuclear reaction:

$$E_{ex} = (1{,}301.4 + 1{,}086.97)/2 = 1{,}194.19 \quad \text{keV} . \qquad (3)$$

With the predicted excitation energy obtained from Eq. 3, and given the small difference between the masses of $^{174}$Lu and $^{175}$Lu, it becomes clear why we could not observe any signs of the $^{174g}$Lu+$d$→$^{176}$Hf nuclear reaction; the excitation energy is insufficient to excite the 10$^{+}$ state (at 1,481.1 keV) for $^{176}$Hf. Now, with the value of the excitation energy, we paid attention to the most intensive cascade of gamma-transitions resulting from the $^{177}$Hf decay. These are identified with the circles and elliptical contours shown in Fig. 3. Also, the gamma-line peak areas are shown in Fig. 4. The only missing gamma-line is at 305.5 keV, and it cannot be resolved, because it is in the shadow of a very intense peak at 306.8 keV. If the 305.5 keV transition were not there, we would not have been able to observe the 249.67 keV gamma-peak, which is clearly present in the instrumental spectrum.

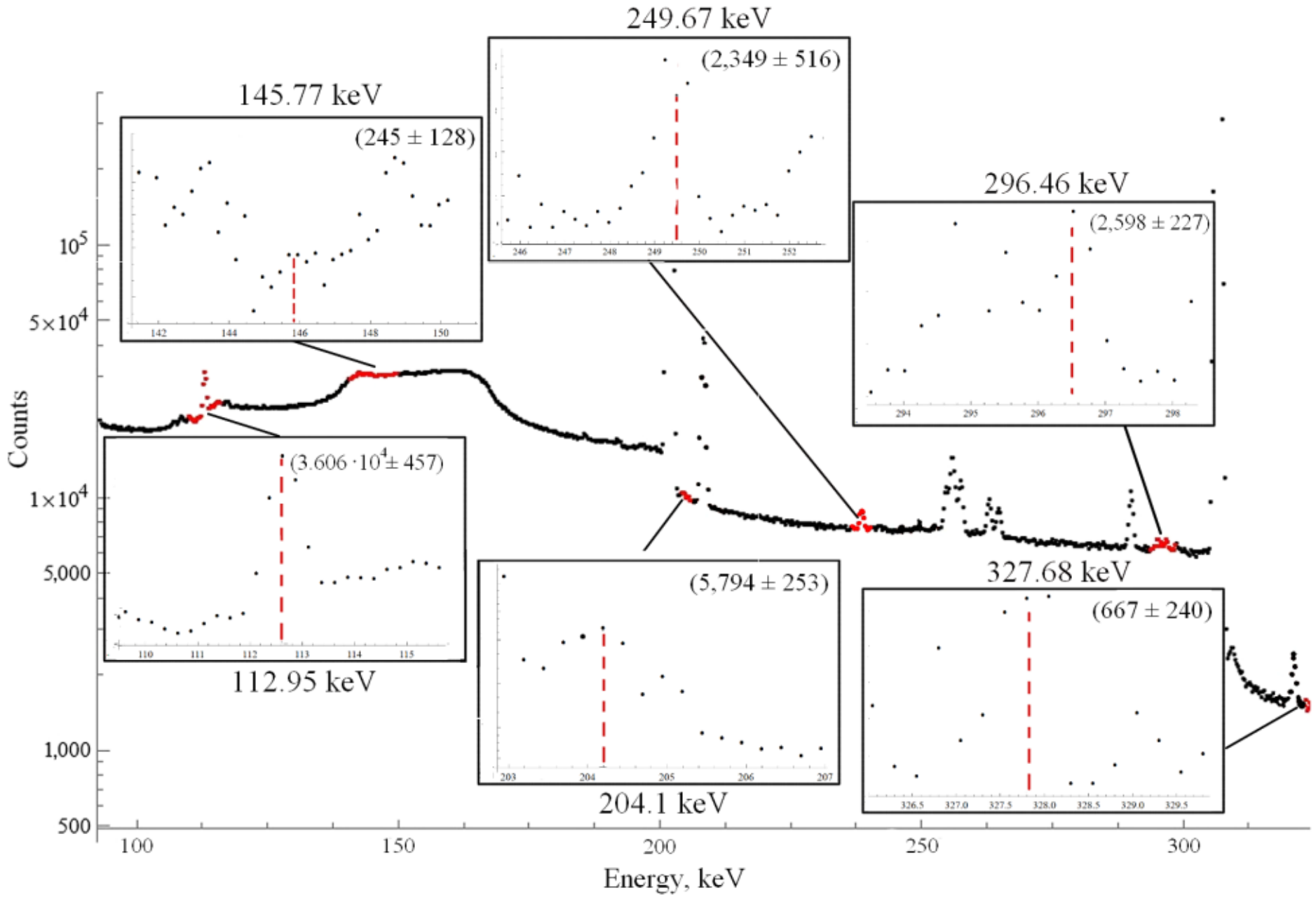

**Fig. 4** Gamma-peaks due to the relaxation of $^{177}$Hf in the spectrum Lu-1 +Lu-2 1$^{st}$

Such a chain implies that a dineutron can also be formed as described in the $^{176}$Lu($n$, $^{2}n$)$^{175}$Lu nuclear reaction. This is true, even though the $^{175}$Lu nucleus was defined as less suitable towards the formation of a nearby dineutron. Hence, after the dineutron's decay, the resulting deuteron fuses with $^{175}$Lu to give rise to the $^{177}$Hf nucleus.

## 4 Discussion

Let us discuss some features of the formation of the bound dineutron in the state 0$^{+}$. In fact, it can be formed near the surface of $^{175}$Lu from an incident neutron and a neutron from $^{176}$Lu. This is somewhat analogous to the dineutron correlation state observed in the case of $^{11}$Li. It is well known that such a configuration consists of a $^{9}$Li core and two neutrons orbiting this core at a distance equal to the total radius of the doubly magic nucleus $^{208}$Pb. In addition, the two neutrons represent a dineutron correlation localized radially in the surface region of $^{11}$Li. In $^{11}$Li, the two neutrons are located quite far from the $^{9}$Li core, and the angle between the neutrons is close to 48 degrees [18]. The $^{11}$Li nucleus exists due to three interactions in the system: between the $^{9}$Li core and each neutron as well as between the neutrons and the wave functions of this system, which forms three parts, overlap. The formation of a dineutron near $^{175}$Lu is possible as a result of these three interactions acting within a potential well at a distance of ~2 fm. Therefore, the existence of a potential pocket for such a system is also highly probable. Moreover, dineutron formation near $^{175}$Lu is possible even with non-zero angular momentum, since the total moment of inertia of an extended dineutron-$^{175}$Lu system can be greater than that of a compact $^{177}$Lu system. Consequently, the rotational energy of the dineutron + $^{175}$Lu nuclear system may be lower than that of $^{177}$Lu. Thus, the formation of the dineutron + $^{175}$Lu system may even be energetically more favorable.

Upon further consideration, the formation of a bound dineutron near the surface of Lu nuclei does not contradict the concept of nuclear molecules [19], first introduced experimentally in 1960 in heavy-ion nuclear physics. Such molecules are a dinuclear system of two nuclei bound together in a quasi-molecular potential, i.e., a configuration of two practically touching nuclei with their potential wells overlaped that retain their individuality before moving apart or fusing into a single nucleus [20]. Moreover, both nuclei can retain their individuality without being recognized by the atom to which they both belong. Once formed, such configurations can exist for their lifetime, showing some analogy with the isomeric state of a nuclear molecule, in which at least one of the two nuclei can exhibit its properties through nuclear decay, and this is what was observed in our experiments as the decay of heavier nuclei ($^{174g}$Lu) in the outgoing channel or fusion into a single nucleus ($^{177}$Hf). Nuclear isomerism of $^{174,175}$Lu plus $^{2}$n dinuclear system is likely caused by the shape (the shape isomerism) and the presence of nuclear levels within the potential well of a heavy nucleus outside its volume as it was shown in [21].

The results described above have demonstrated a marked discrepancy between the measured cross-section for the $^{176}$Lu($n$, $\gamma$)$^{177m,g}$Lu reaction and the theoretical predictions provided by well tested nuclear reaction models and also by databases such as TALYS-1.96, Empire 3.2.3, TENDL, and ENDF [8]. Despite thorough consideration of scattered neutron contributions, the observed enhancement remained unexplained by conventional means. This necessitated the suggestion of alternative nuclear pathways, including those reaction channels beyond the well-known and established standard formalism. A potential explanatory hypothesis involves the presence of a previously unconsidered nuclear mechanism involving the formation of a bound dineutron.

Specifically, the chain nuclear reaction mechanism:

$$^{176}Lu(n,{}^{2}n)^{175}Lu \xrightarrow{^{2}_{0}n\ decay} {}^{175}Lu(d,\gamma)^{177}Hf$$

suggests that a singlet state bound dineutron, within the potential well of $^{175}$Lu, undergoes $\beta^{-}$ - decay [11] to yield a deuteron [5], which subsequently fuses with the host $^{175}$Lu nucleus. This mechanism ultimately populates excited states of the stable $^{177}$Hf, whose prompt gamma emissions may have been misattributed to the decay of $^{177m,g}$Lu, thus artificially inflating the measured cross-section.

Then in the absence of a sufficient number of other particles and isotopes that are relevant for the $^{176}$Lu($n$, $\gamma$)$^{177m,g}$Lu reaction to be a reality, we have to admit that there is a reaction involving dineutron formation; wherein the dineutron is hosted by the $^{175}$Lu nucleus. Decay of the dineutron is followed by fusion with the product nuclei. The dineutron itself exists within the potential well of the nucleus, as well as its decay products. That is, they are within the range of the strong interaction, i.e. a few fm. Therefore, the decay products, in particular, the deuteron can fuse with the $^{175}$Lu nucleus. This is true, because the strong force overwhelms any Coulombic repulsion. For this process to take place, the deuteron must possess some kinetic energy and garner acceleration, causing excitation of the residual nucleus.

Similarly, the analysis of the $^{175}$Lu($n$, 2$n$)$^{174g}$Lu reaction, with its measured cross-sections that exceed the TENDL estimates by over 15-25%, further reinforces the suggestion that some hidden nuclear reaction channels are open. This excess might be partially attributed to the $^{176}$Lu($n$, 3$n$)$^{174g}$Lu nuclear reaction, for which only theoretical cross section estimations exist. The lack of the experimental validation of the magnitude of the deviation, draws attention to the importance of considering the more exotic $^{176}$Lu($n$, $^{2}n$+$n$)$^{174g}$Lu nuclear reaction channel. In addition, if we look at Fig. 7 in reference [12], along with the results of the measurements and evaluations for the $^{175}$Lu($n$, 2$n$)$^{174g}$Lu nuclear reaction, then we see that the majority of the available measured cross-sections are greater than the theoretically predicted cross-section. Both the cross-section enhancements, at 14-MeV neutron energy, need additional provident and confirmation for the same effect in other energy ranges. Therefore, we considered our results of cross-section measurement for lower neutron energies ~ 6 MeV [5] in order to prove the hypothesis about the formation of bound dineutrons at higher energies of the impinging neutrons.

The anomalous behavior of $^{176g}$Lu nuclei, under fast neutron irradiation, was also demonstrated by the post-irradiation "half-life" of approximately 62 days obtained from Eq. 1. This result contradicts the known half-life of this isotope ($T_{1/2}$=3.76×10$^{10}$ y). This deviation does, however, refer to a characteristic nuclear fuel depletion in reactor physics, wherein neutron capture results in nuclide burnup (e.g. transmutation). The calculated effective cross-section of ~3.9×10$^{11}$ b, obtained from Eq. 2, while clearly non-physical in traditional nuclear reaction frameworks, underscores the presence of a potent removal mechanism. The mechanism is most plausibly neutron-induced transmutation, followed by nuclear fusion via intermediate light clusters, such as deuterons. Such a huge reaction cross-section estimate may be the evidence of the following prediction in [21]: “This phenomenon can be observed in nuclear reactions, where it manifests itself as a resonant input channel.” It means that some neutron, in the energy range of 5.51 - 6.00 (5.76 average) MeV [5], does correspond to that resonant neutron energy.

In addition, from Table 2, one can stress that 88.3 and 201.8 keV gamma-photons are due to the decay of depleted $^{176}$Lu (Fig. 2), and they are even lower than expected. In this study, we intentionally did not consider the background aspects in detail, in particular the 88.3 keV gamma-peak, which overlaps with other peaks [5]. Should we do so, the difference between count rates would become even greater. A reasonable explanation for this anomaly could be, after fast neutron irradiation of the Lu is complete, that a higher number of Auger and conversion electrons occur per unit time [15]. However, the reason for that occurrence remains unclear. In addition, to make sure the detection efficiency did not deteriorate, we checked it before the first spectrum measurement for (Lu-1 + Lu-2 BS) [5] and after the last one for (Lu-1 +Lu-2 1$^{st}$); no changes were noticed.

Another interesting finding is the fact that fusion between a heavier nucleus of $^{175}$Lu and the deuteron takes place at ambient conditions. This result does not violate anything, but it proves the theoretical prediction reported in [21], according to which the distance is a few fm between the two atomic nuclei of very different masses. The lighter nuclei must exist in the potential well of the heavier atomic nucleus before such a synthesis takes place. Moreover, with the estimate of the excitation energy ~1.2 MeV from Eq. 3 and the Q value of ~10.85 MeV for the $^{175}$Lu($d$, $\gamma$)$^{177}$Hf nuclear reaction [3], we find that a 9.65 MeV kinetic energy transfer from $^{177}$Hf to the sample material is possible. This 9.65 MeV energy estimate can be even greater due to some extra kinetic energy of the deuteron as discussed above. $^{177}$Hf is a stable nucleus, and this exoenergetic mechanism is not accompanied by the production of radioactive waste. Even more, the highly effective transmutation of long-lived $^{176g}$Lu is experimentally demonstrated.

## 5 Conclusions and outlook

We analyzed the results of cross-section measurements for the nuclear reactions involving Lu isotopes under fast neutron irradiation, with special focus on the $^{176}$Lu($n$, $\gamma$)$^{177m,g}$Lu, $^{175}$Lu($n$, $^2n$)$^{174g}$Lu, $^{176}$Lu($n$, $^2n$)$^{175}$Lu and $^{174,175}$Lu($d$, $\gamma$)$^{176,177}$Hf channels. The measured ($n$, $\gamma$) cross-sections significantly exceed the values that can be obtained using the codes TALYS and EMPIRE or taken from the TENDL, and ENDF nuclear data libraries, indicating the potential contribution from alternative reaction mechanisms. One such proposed mechanism is the formation of a bound dineutron in the ($n$, $^2n$) nuclear reaction, which undergoes $\beta^-$ -decay to form a deuteron. The resulting fusion of a deuteron with Lu nuclei can lead to Hf isotopes, particularly $^{177}$Hf, imitating the gamma-signature of $^{177m,g}$Lu decay. It then becomes possible to study these nuclear properties to clearly identify such a unique nuclear structure and determine its extremely important characteristics before the two nuclei with atomic masses A and 2 separate or merge into an A+2 nucleus.

Additional experimental evidence shows consistent suppression of the $^{176g}$Lu decay gamma lines after relatively short neutron irradiation, supporting a burnup model with a dramatically decreased half-life of $^{176}$Lu (from $3.76\times10^{10}$ y to 62.2 d) at certain time interval and an anomalously large effective cross-section ($\approx 3.9\times10^{11}$ b). Such a large cross-section cannot be explained by any of the known reaction mechanism, given that, according to [4], the cross-section for the $^{176}$Lu($n$, $tot$) nuclear reaction in this incident neutron energy range is ~ 2.5 b.

Gamma-spectrometry revealed peak patterns consistent with the fusion-produced excited states of $^{177}$Hf that may be validating the dineutron fusion and/or deuteron fusion with the residual nucleus hypothesis. For example, the dineutron as a separate nucleus decays to a deuteron located within the nuclear force range of the residual nucleus, allowing low-energy fusion without the need for thermonuclear conditions. Observed gamma-transitions and the absence of high-energy gamma-lines in $^{176,177}$Hf confirm that excitation energies from the corresponding reactions on $^{175,176}$Lu are below the energies of levels required to populate the specific $^{176,177}$Hf nuclear states. In other words, we have experimentally proven that the excitation energy of $^{177}$Hf, produced via the fusion of $^{175}$Lu, with the deuteron is about 1.2 MeV under ambient temperatures. Then this might be an interesting case when a long-lived radionuclide can be converted into an energy source.

Experiments on the irradiation of natural lutetium with fast neutrons have produced reaction products and isotope depletion effects that cannot be explained within the framework of known nuclear reaction channels. The observations can be interpreted consistently by assuming the formation of a bound dineutron in certain resonant ($n$, $^2n$) reactions with the above-described outcomes. Consequently, our experimental neutron energy, $5.76^{+0.24}_{-0.25}$ MeV, allows us to claim that this same resonant energy is needed to populate the dineutron level in the $^{176}$Lu($n$,$^2n$)$^{175}$Lu nuclear reaction.

Getting back to the beginning of this narration, we can reconsider some of the published results in [1,2] as activation cross-sections not of some radio-isotope and naturally occurring heavy element, but rather of a bound dineutron, and the emission of one more neutron in the outgoing channel. The destiny of this neutron is very interesting, since it either emits or is captured at another Migdal's single-particle level. The capture can only occur when the dineutron, or its daughter particle (the deuteron) is within the potential well of the residual nucleus. As far as this phenomenon takes place, a bound state of three neutrons, or the 'tri-neutron' may occur according to the implicit statement in [21].

In summary, the presence of unobserved dineutron escape and fusion-like channels explains the enhanced gamma activity, depletion of nuclei, and calculated cross-sections. This further introduces a novel low-energy reaction and fusion path under neutron irradiation conditions, with implications for nuclear structure, reaction modeling, the study of dark matter [22], and practical applications [23]. These results are in correspondence with the expectations of new Nuclear Structure and Reaction Dynamics trends, in particular about the mechanisms behind known nuclear reactions [24].

**Acknowledgements** The authors are grateful to academician Prof. V. Yu. Denisov of Kyiv Institute for Nuclear Research of National Academy of Sciences of Ukraine and prof. C. Stevenson of Illinois State University for fruitful discussions and more precise explanations of the results obtained.

**Author Contributions** All authors contributed to the study conception and design. Methodology proposition, investigation performance were made by Ihor Kadenko, András Fenyvesi and Kateryna Okopna. Data curation was performed by Barna Biró, András Fenyvesi, Vladyslav Morozyuk, Kateryna Okopna and Nadiia Sakhno. Resources searching, supervision and project administration, funding acquisition were executed by Ihor Kadenko and András Fenyvesi. Formal analysis was made by Barna Biró and Vladyslav Morozyuk. Simulation was under responsibility of András Fenyvesi and Vladyslav Morozyuk. Validation was executed by Barna Biró, Kateryna Okopna and Nadiia Sakhno. Visualization was implemented by Barna Biró, Vladyslav Morozyuk, Kateryna Okopna and Nadiia Sakhno. Original draft of manuscript was written by Ihor Kadenko and Kateryna Okopna, all authors commented on and edited previous versions of the manuscript. All authors read and approved the final manuscript.

**Declarations Conflict of interest** All authors declare that there are no conflict of interest.

## References

1. S. M. Qaim, M. Ejaz, Half-lives and activation cross-sections of some radio-isotopes of iodine, tellurium and antimony formed in the interactions of iodine with 14-7 MeV neutrons, J. Inorg. Nucl. Chem. **30**, 2577-2581 (1968). https://doi.org/10.1016/0022-1902(68)80382-3
2. A.K. Hankla, R.W. Fink, J.H. Hamilton, Neutron activation cross-sections at 14.4 MeV for some naturally occurring heavy elements in the region 76 ≤ Z ≤ 82, Nucl. Phys. A **180**, 157-176 (1972). https://doi.org/10.1016/0375-9474(72)90162-5
3. https://www.nndc.bnl.gov/qcalc/
4. A.J. Koning, D. Rochman, J.Ch. Sublet, TALYS-based evaluated nuclear data library, TENDL-2023, https://tendl.web.psi.ch/tendl_2023/tendl2023.html
5. I. Kadenko, B. Biró, A. Fenyvesi et al., Formation of bound dineutrons in the $^{175}$Lu (n, $^{2}$n) $^{174g}$Lu nuclear reaction and its cross-section, Phys. Lett. B **859**, 139100 (2024). https://doi.org/10.1016/j.physletb.2024.139100
6. I. Kadenko, Possible observation of the dineutron in the $^{159}$Tb (n, $^{2}$n)$^{158g}$Tb nuclear reaction, Europhys. Lett. **114**, 42001 (2016). DOI:10.1209/0295-5075/114/42001
7. I. Kadenko, B. Biro, A. Fenyvesi, Statistically significant observation of and cross-sections for a new nuclear reaction channel on $^{197}$Au with bound dineutron escape, Europhys. Lett. **131**, 52001 (2020). DOI 10.1209/0295-5075/131/52001
8. A.M. Savrasov, O.M. Gorbachenko, V.M. Morozyuk, et al., Investigation of fast neutron reactions in natural lutetium, Nucl. Phys. A **1041**, 122788 (2024). https://doi.org/10.1016/j.nuclphysa.2023.122788
9. N. Dzysiuk. I. Kadenko, A.J. Koning et al., Cross sections for fast-neutron interaction with Lu, Tb, and Ta isotopes, Phys. Rev. C **81**, 014610 (2010). https://doi.org/10.1103/PhysRevC.81.014610
10. N. Dzysiuk, A. Kadenko, I. Kadenko et al., Measurement and systematic study of (n,$x$) cross sections for dysprosium (Dy), erbium (Er), and ytterbium (Yb) isotopes at 14.7 MeV neutron energy, Phys. Rev. C **86**, 034609 (2012). https://doi.org/10.1103/PhysRevC.86.034609
11. Kadenko I.M., Sakhno N.V., Gorbachenko O.M. et al., Delayed $^{160}$Tb radioactivity build-up due to $^{159}$Tb(n,$^{2}$n) nuclear reaction products transformation and subsequent fusion, Nucl. Phys. A **1030**, 122575 (2023). https://doi.org/10.1016/j.nuclphysa.2022.122575
12. Y. Song et al., F. Zhou, Y. Li et al., Cross sections for 14 MeV neutron interaction with lutetium isotopes and their theoretical excitation functions, Chinese Phys. C **48**, 054104 (2024). DOI:10.1088/1674-1137/ad2b55
13. [dataset] Experimental Nuclear Reaction Data (EXFOR), Database Version of 2024-04-08, 2024.
14. N. Dzysiuk, I.M. Kadenko, O.O. Prykhodko, Candidate-nuclei for observation of a bound dineutron. Part I: The (n,$^{2}$n) nuclear reaction, Nucl. Phys. A **1041,** 122767 (2024). https://doi.org/10.1016/j.nuclphysa.2023.122767
15. [dataset] NuDat 3 Data Base, National Nuclear Data Center (NNDC), Brookhaven National Laboratory, Upton, NY, USA. https://www.nndc.bnl.gov/nudat3/ (last accessed 03.09.2025)
16. R.A. Knief, *Nuclear Engineering: theory and technology of commercial nuclear power*, second ed., (ANS American Nuclear Society, Inc., 2014).
17. I.M. Kadenko, N.V. Sakhno, Possible LENR observation due to dineutron formation as a product of the $^{159}$Tb(n,$^{2}$n)$^{158}$Tb nuclear reaction, Acta Physica Polonica B **51** (1), 83-89 (2020).

https://doi.org/10.5506/APhysPolB.51.83

18. Y. Kubota, A. Corsi, G. Authelet et al., Surface Localization of the Dineutron in $^{11}$Li, Phys. Rev. Lett. **125**, 252501 (2020) https://doi.org/10.1103/PhysRevLett.125.252501
19. W. Greiner, J.Y. Park, W. Scheid. *Nuclear molecules*, World scientific publishing, 1995, 465 p.
20. G.G. Adamian, N.V. Antonenko, R.V. Jolos etc. Intern. J. of Modern Phys. E **16** (4), 1021-1031 (2007). DOI: 10.1142/S0218301307006472
21. A.B. Migdal, Two interactive particles in a potential well, Yad. Fiz. **16**, 427 (1972) [Two interactive particles in a potential well, Sov. J. Nucl. Phys. **16**, 238-241 (1973)] (**Russian journal reference with English journal translation**)
22. Y. Hao, Zh. Chen, Dark photons from dineutron decays in neutron stars, Phys. Rev. D **112**, 075017 (2025). https://doi.org/10.1103/sqn7-fd3z
23. Ch.D. Stevenson, J.P. Davis, Quantum field stabilization of the di-neutron enabling low energy deuterium fusion, Int. J. Hydrog. Energy **61**, 1-5 (2024). https://doi.org/10.1016/j.ijhydene.2024.02.116
24. Perspectives for European Nuclear Physics, The NuPECC Long Range Plan 2024, https://www.nupecc.org/lrp2024/Documents/nupecc_lrp2024.pdf